\documentclass[journal]{IEEEtran}

\usepackage{graphics,amsmath,amsfonts,amscd,latexsym,enumerate,multirow,epsfig}

\usepackage{array}

\begin{document}

\def\<{\langle}
\def\>{\rangle}
\def\tr{ {\rm{Tr \,}}}
\def\supp{ {\rm{supp \,}}}
\def\dim{ {\rm{dim \,}}}
\def\oti{{\otimes}}
\def\bra#1{{\langle #1 |  }}
\def\lb{ \left[ }
\def\rb{ \right]  }
\def\tilde{\widetilde}
\def\bar{\overline}
\def\*{\star}

\def\({\left(}      \def\BL{\Bigr(}
\def\){\right)}     \def\BR{\Bigr)}
    \def\BBL{\lb}
    \def\BBR{\rb}
%

\def\E{\mathbb{E}}
\def\1{I}

\def\bb{{\bar{b} }}
\def\ab{{\bar{a} }}
\def\zb{{\bar{z} }}
\def\zbar{{\bar{z} }}
\def\frac#1#2{{#1 \over #2}}
\def\inv#1{{1 \over #1}}
\def\half{{1 \over 2}}
\def\d{\partial}
\def\der#1{{\partial \over \partial #1}}
\def\dd#1#2{{\partial #1 \over \partial #2}}
\def\vev#1{\langle #1 \rangle}
\def\ket#1{ | #1 \rangle}
\def\bra#1{\langle #1 |}
\def\rvac{\hbox{$\vert 0\rangle$}}
\def\lvac{\hbox{$\langle 0 \vert $}}
\def\2pi{\hbox{$2\pi i$}}
\def\e#1{{\rm e}^{^{\textstyle #1}}}
\def\grad#1{\,\nabla\!_{{#1}}\,}
\def\dsl{\raise.15ex\hbox{/}\kern-.57em\partial}
\def\Dsl{\,\raise.15ex\hbox{/}\mkern-.13.5mu D}
\def\b#1{\mathbf{#1}}
\newcommand{\proj}[1]{\ket{#1}\bra{#1}}
%
%
\def\th{\theta}     \def\Th{\Theta}
\def\ga{\gamma}     \def\Ga{\Gamma}
\def\be{\beta}
\def\al{\alpha}
\def\ep{\epsilon}
\def\vep{\varepsilon}
\def\la{\lambda}    \def\La{\Lambda}
\def\de{\delta}     \def\De{\Delta}
\def\om{\omega}     \def\Om{\Omega}
\def\sig{\sigma}    \def\Sig{\Sigma}
\def\vphi{\varphi}
%
%
\def\CA{{\cal A}}   \def\CB{{\cal B}}   \def\CC{{\cal C}}
\def\CD{{\cal D}}   \def\CE{{\cal E}}   \def\CF{{\cal F}}
\def\CG{{\cal G}}   \def\CH{{\cal H}}   \def\CI{{\cal J}}
\def\CJ{{\cal J}}   \def\CK{{\cal K}}   \def\CL{{\cal L}}
\def\CM{{\cal M}}   \def\CN{{\cal N}}   \def\CO{{\cal O}}
\def\CP{{\cal P}}   \def\CQ{{\cal Q}}   \def\CR{{\cal R}}
\def\CS{{\cal S}}   \def\CT{{\cal T}}   \def\CU{{\cal U}}
\def\CV{{\cal V}}   \def\CW{{\cal W}}   \def\CX{{\cal X}}
\def\CY{{\cal Y}}   \def\CZ{{\cal Z}}

\def\bea{\begin {eqnarray*}}
\def\eea{\end {eqnarray*}}

\def\rvac{\hbox{$\vert 0\rangle$}}
\def\lvac{\hbox{$\langle 0 \vert $}}
\def\comm#1#2{ \BBL\ #1\ ,\ #2 \BBR }
\def\2pi{\hbox{$2\pi i$}}
\def\e#1{{\rm e}^{^{\textstyle #1}}}
\def\grad#1{\,\nabla\!_{{#1}}\,}
\def\dsl{\raise.15ex\hbox{/}\kern-.57em\partial}
\def\Dsl{\,\raise.15ex\hbox{/}\mkern-.13.5mu D}
\def\beq{\begin {equation}}
\def\eeq{\end {equation}}
\def\to{\rightarrow}
\def\To{\Rightarrow}
\def\mtx#1{|#1\rangle\langle#1|}
\def\FT{\Pi_{\rho,\delta}^{n}}
\def\CFT{\check{\Pi}_{\rho,\delta}^{n}}
\def\FTN{\Pi_{\CN(\rho),\delta}^{n}}
\def\CFTN{\check{\Pi}_{\CN(\rho),\delta}^{n}}
\def\pt{p_\alpha}
\def\CNn{\CN^{\otimes n}}
\def\CMn{\CM^{\otimes n}}
\def\rhon{\rho^{\otimes n}}
\def\on{{\otimes n}}
\def\U{U_{s^a}}
\def\S{\sigma_{s^a}}
\def\P{\Pi_{s^a}}
\def\pit{\pi^n_{\alpha}}
\def\Ain{\varphi_{\rho_1}^\on}
\def\Bin{\varphi_{\rho_2}^\on}
\newtheorem{theorem}{Theorem}
\newtheorem{acknowledgement}[theorem]{Acknowledgement}
\newtheorem{algorithm}[theorem]{Algorithm}
\newtheorem{axiom}[theorem]{Axiom}
\newtheorem{claim}[theorem]{Claim}
\newtheorem{conclusion}[theorem]{Conclusion}
\newtheorem{condition}[theorem]{Condition}
\newtheorem{conjecture}[theorem]{Conjecture}
\newtheorem{corollary}[theorem]{Corollary}
\newtheorem{criterion}[theorem]{Criterion}
\newtheorem{definition}[theorem]{Definition}
\newtheorem{example}[theorem]{Example}
\newtheorem{exercise}[theorem]{Exercise}
\newtheorem{lemma}[theorem]{Lemma}
\newtheorem{notation}[theorem]{Notation}
\newtheorem{problem}[theorem]{Problem}
\newtheorem{proposition}[theorem]{Proposition}
\newtheorem{remark}[theorem]{Remark}
\newtheorem{solution}[theorem]{Solution}
\newtheorem{summary}[theorem]{Summary}

%
\title{Entanglement-Assisted Capacity of Quantum Multiple Access Channels}
%
%
%

\author{Min-Hsiu Hsieh, Igor Devetak and Andreas Winter%
\thanks{Manuscript received February 21, 2006; revised January 29, 2008. The work of M.-H. Hsieh and I. Devetak was supported by the National Science Foundation under NSF 05-501 Grant 0524811. The work of A.~Winter was supported by the European Union (EU) under Grant RESQ, no.~IST-2001-37559 and the United Kingdom Engineering and Physical Sciences Research Council's ``QIP IRC''.}
\thanks{M.-H. Hsieh is with the Electrical Engineering Department, University
of Southern California, Los Angeles, CA 90089 USA (e-mail: minhsiuh@
usc.edu).}
\thanks{I. Devetak is with University of Southern Califonia, USC Viterbi School of
Engineering, Los Angeles CA 90089-2565 USA (e-mail: devetak@usc.edu).}
\thanks{A. Winter is with Department of Mathematics, University of Bristol, Bristol
BS8 1TW, U.K. (e-mail: A.J.Winter@bristol.ac.uk).}}


%


\maketitle

\begin{abstract}
We find a regularized formula for the entanglement-assisted (EA)
capacity  region for quantum multiple access channels (QMAC). We
illustrate the capacity region calculation with the example of the
collective phase-flip channel which admits a single-letter
characterization. On the way, we provide a first-principles proof of
the EA coding theorem based on a packing argument. We observe that
the Holevo-Schumacher-Westmoreland theorem may be obtained from a
modification of our EA protocol. We remark on the existence of a
family hierarchy of protocols for multiparty scenarios with a single
receiver, in analogy to the two-party case. In this way, we relate
several previous results regarding QMACs.
\end{abstract}


\section{Introduction}
\IEEEPARstart{S}{hannon's} classical channel capacity theorem is one
of the central results in classical information theory
\cite{Shannon48}. A single-sender channel is defined by the triple
$(\CX, p(y|x),\CY)$ where the sets $\CX$ and $\CY$ represent the
input and output alphabets, respectively, and the conditional
distribution $p(y|x)$ defines the probability of the output being
$y$ given that the input was $x$. The capacity $C$ of the channel,
the maximum rate at which classical information can be transmitted
through the channel, is given in terms of the mutual information
$I(X;Y) = H(X) + H(Y) - H(XY)$, (here the entropy of a random
variable $X$ with probability distribution $p(x)$ is given by $H(X)
= -\sum_{x \in \CX} p(x) \log p(x)$):
\begin{equation}
\label{css} C =\max_{p(x)} I(X;Y)
\end{equation}
where the joint distribution of $XY$ is $p(x)p(y|x)$.

The classical multiple-access (MAC) channel
$(\CX\times\CY,p(z|x,y),\CZ)$ is a channel with two senders and one
receiver. Now $\CX$ and $\CY$ are the input alphabets of the first
and second sender, respectively. A general overview of MACs can be
found in \cite{CT91,CK81}. The capacity problem now involves finding
the region of achievable transmission rates $R_1$ and $R_2$ for the
two senders. The classical capacity region of a MAC was found
independently by Ahlswede \cite{Ahlswede71} and Liao \cite{Liao72}.
It is given by the closure of the convex hull of all $(R_1,R_2)$
satisfying
\begin{equation}
\begin{split}
\label{cmac1}
R_1 &\leq I(X;Z|Y)   \\
R_2 &\leq I(Y;Z|X)    \\
R_1+R_2 &\leq I(XY;Z)
\end{split}
\end{equation}
for some product distribution $p(x)p(y)$ on $\CX\times\CY$. Here the
joint distribution of $X Y Z$ is $p(x)p(y)p(z|x ,y)$, and the
conditional mutual information is defined as $I(X;Z|Y) = I(X;YZ) -
I(X;Y)$.

The theory of quantum channels is richer, and includes several
distinct capacities depending on the type of information one is
trying to send and the additional resources one can use. A quantum
channel $\CN$ is modeled as a cptp (completely positive and trace
preserving) map. The capacity $C(\CN)$ of a quantum channel is
defined to be the maximum rate at which classical information can be
sent through the quantum channel $\CN$. This capacity was proved
independently by Holevo \cite{Hol98} and Schumacher and Westmoreland
\cite{SW97}. The capacity $Q(\CN)$ is defined to be the maximum rate
at which quantum information can be sent through the quantum channel
$\CN$, and a formula for it was proven in
\cite{Lloyd96,Shor02,Devetak03}.

Entanglement shared between sender and receiver is a useful resource
that generically increases channel capacity. The
entanglement-assisted classical capacity $C_E(\CN)$ is the maximum
rate at which classical information can be transmitted through the
quantum channel $\CN$ if the sender and receiver have access to
unlimited entanglement. A remarkably simple formula for this
capacity was found in \cite{BSST01, Hol01a}, to be formally
identical to (\ref{css}), with classical mutual information replaced
by the quantum mutual information between quantum systems $A$
and $B$
\begin{equation}
\label{bsst} C_E(\CN) = \max_{\rho} I(A;B).
\end{equation}
The maximization is performed over the sender's input state $\rho$,
and the quantum mutual information $I(A;B)$ is defined with respect
to the purification of $\rho$ after half of it has passed through
the channel $\CN$. The system $A$ is the half remaining on the
sender's side, and $B$ is the channel output system.  Formal
definitions of these concepts will be given in Section \ref{II}.

A quantum multiple access  channel (QMAC) $\CM$ is a cptp map with
two senders and one receiver. Each sender can transmit either
classical or quantum information through the channel $\CM$. The
classical-classical capacity region $C(\CM)$ for the case in which
both senders transmit classical information through QMAC $\CM$ was
found by Winter \cite{Winter01CQ}. Later on, the classical-quantum
capacity region $CQ(\CM)$ (where one sender is sending classical,
and the other quantum information), and the quantum-quantum channel
capacity region $Q(\CM)$ were found in \cite{Yard05, HOW05}.

In this work we consider the entanglement-assisted
classical-classical capacity region $C_E(\CM)$ of a QMAC $\CM$. In
other words, both senders share unlimited entanglement with the
receiver and both are sending classical information. We will show it
to be the \emph{regularized} closure of the set of all the
achievable rate pairs $(R_1,R_2)$ satisfying
\begin{equation}
\begin{split}
\label{qmac}
R_1 &\leq I(A;C|B)   \\
R_2 &\leq I(B;C|A)    \\
R_1+R_2 &\leq I(AB;C)
\end{split}
\end{equation}
for some choice of a product input state $\rho_1 \otimes \rho_2$ for
the two senders. The quantum entropic quantities are defined with
respect to the product of purifications of $\rho_1$ and $\rho_2$,
after half of it has passed through the channel $\CM$. The systems
$A$ and $B$ are the parts remaining on the senders' sides, and $C$
is the channel output system. A precise statement of the result is
given in Theorem 2. The expression (\ref{qmac}) thus parallels
(\ref{cmac1}) with the classical mutual information replaced by its
quantum counterpart. While our formula does  not allow $C_E(\CM)$ to
be efficiently computed in general, we exhibit a non-trivial example
for we can compute $C_E(\CM)$ in closed form.

We also provide a new proof of the direct coding theorem for the
single-sender entanglement-assisted channel capacity. Our proof is
important and necessary in the following sense. First, our proof
uses packing lemma that comes from the idea of typical subspaces,
which is directly analog to the idea of typical sets Shannon uses to
prove the direct coding theorem of single-user channel capacity. The
previous proof in \cite{BSST01,Hol01a} is less trivial in the sense
that it is based on the Holevo-Schumacher-Westmoreland (HSW) theorem
\cite{Hol98,SW97}, which uses the conditional typical subspaces. Our
proof demonstrates our growing understanding of quantum information
theory. We believe that our method of proof will not only become a
powerful tool but also will find many applications in quantum
information theory. Second, our proof provides new properties that
can be used to prove the multiparty generalization. These new
properties do not exist in the previous proofs. Finally, we show
that the HSW theorem is a special case of the two-party
entanglement-assisted capacity theorem.

The paper is organized as follows. Section \ref{II} contains the
relevant background material. This includes  notational conventions,
definitions of the method of types, frequency typical sequences and
subspaces, and useful lemmas. Section \ref{III} contains statements
and proofs of our main results. In section \ref{IV} we compute the
capacity region of the collective phase-flip multiple access channel
which admits a single-letter expression. In section \ref{V} we first
rewrite our results in the resource inequality framework, from which
we recover previously known coding theorems for QMACs. In section
\ref{VI} we conclude by pointing out the open question regarding the
single-letter expression for our entanglement-assisted capacity
region of quantum multiple access channels. We also give a
conjecture on the entanglement-assisted channel capacity with more
than two inputs.

\section{Background}
\label{II} Each quantum system is completely described by the state
vector which is a unit vector in Hilbert space $\CH$. An alternative
way to describe a quantum system is by density operator
$\rho:\CH\to\CH$, where $\rho$ has trace equal to one and is a
positive operator. If $\rho$ belongs to a quantum system $A$ we may
denote it by $\rho^A$. When it is clear from contexts, we will omit
the superscript letter that represents the holder of the quantum
system. We always use $\pi$ to denote the maximally mixed state
$\pi= (|\CH|)^{-1} I$ where $|\CH|$ represents the dimension of
$\CH$. Given a state $\rho^A$ whose spectral decomposition is
$\sum_{i}p_i\proj{i}$, the purification of such state is obtained by
introducing a reference system $R$ such that the purified state
$\ket{\psi}^{AR}=\sum_{i}\sqrt{p_i}\ket{i}^A\ket{i}^R$. We write the
density operator of a pure state $\ket{\psi}$ as
$\psi\equiv\ket{\psi}\bra{\psi}$.

Saying that $\CN: A \rightarrow B$ is a quantum channel, we really
mean that $\CN:\CB(\CH_A) \to \CB(\CH_B)$ is a cptp (completely
positive trace preserving) map, where $\CB(\CH)$ represents the set
of bounded linear operators in $\CH$. It may be modeled by an
isometry $U_\CN: A \rightarrow BE$ with a larger target space $BE$,
followed by tracing out the ``environment'' system $E$. $U_\CN$ is
known as the Stinespring dilation \cite{stinespring55} of $\CN$. We
will often write $U_\CN(\rho)$ for $U_\CN \, \rho \, U^\dagger_\CN$.

A quantum instrument \cite{Davies70,Ozawa84} $\mathbf{D} =
\{\CD_m\}_{m\in[\mu]}$, $[\mu]:=\{1,2,\cdots,\mu\}$, is a set of cp
(completely positive) maps $\CD_m$,
$$
\CD_m: \rho \to \sum_k A_{km} \rho A^\dagger_{km}.
$$
The sum of the cp maps $\CD = \sum_{m \in [\mu]} \CD_m$ is trace
preserving, and $\sum_{km} A^\dagger_{km} A_{km} = I$. The
instrument has one quantum input and two outputs, classical and
quantum. The probability of classical outcome $m$ and corresponding
quantum output $\CD_m(\rho)/ (\tr \CD_m(\rho))$ is  $\tr
\CD_m(\rho)$. Ignoring the classical output reduces the instrument
to the quantum map $\CD$. Ignoring the quantum output reduces the
instrument to the set of POVMs (positive operator valued measure)
$\{\Lambda_m\}$ with $\Lambda_m = \sum_k A^\dagger_{km} A_{km}$.

The trace distance is defined as the trace norm of the difference
between the two states
$$\|\sigma-\rho\|=\tr{\sqrt{(\sigma-\rho)^2}} =
\max_{-\1 \leq \Lambda \leq \1} \tr [\Lambda (\sigma-\rho)].$$

\medskip

The method of types is a standard technique of classical information
theory. Denote by $x^n$ a sequence $x_1x_2\dots x_n$, where each
$x_i$ belongs to the finite set $\CX$. Denote by $|\CX|$ the
cardinality of $\CX$. Denote by $N(a|x^n)$ the number of occurrences
of the symbol $a\in\CX$ in the sequence $x^n$. The \emph{type}
$t^{x^n}$ of a sequence $x^n$ is a probability vector whose elements
$t^{x^n}_{a} = \frac{N(a|x^n)}{n}$. Denote the set of sequences of
type $t$ by
$$\CT^n_t=\{x^n\in \CX^n: t^{x^n}=t\}.$$
For the probability distribution $p$ on the set $\CX$ and $\delta
>0$, let $\tau_\delta=\{t:\forall a \in \CX,\ |t_a-p_a|\leq \delta\}$.
Define the set of $\delta$-typical sequences of length $n$ as
\begin{equation}
\begin{split}
\CT^{n}_{p,\delta}&=\underset{t \in \tau_\delta}{\bigcup} \CT^n_t   \\
&=\{x^n:\forall a\in\CX,\ |t^{x^n}_a-p_a|\leq \delta\}.
\end{split}
\end{equation}
Define the probability distribution $p^n$ on $\CX^n$ to be the
tensor power of $p$. The sequence $x^n$ is drawn from $p^n$ if and
only if each letter $x_i$ is drawn independently from $p$. Typical
sequences enjoy many useful properties. Let $H(p) = - \sum_x p_x
\log p_x$ be the Shannon entropy of $p$. For any $\ep,\delta>0$, and
all sufficiently large $n$ for which
\begin{equation}
p^n(\CT_{p,\delta}^{n})  \geq  1-\ep \label{cc1}
\end{equation}
\begin{equation}
2^{-n\lb H(p)+c\delta \rb}  \leq  p^n(x^n)  \leq   2^{-n\lb
H(p)-c\delta\rb},  \,\,\, \forall x^n \in \CT_{p,\delta}^{n}
\label{cc2}
\end{equation}
\begin{equation}
 |\CT_{p,\delta}^{n}| \leq   2^{n\lb H(p)+c\delta\rb}
\label{cc3}
\end{equation}
for some constant $c$ (see \cite{CT91} for proofs). For $t\in
\tau_\delta$ and for sufficiently large $n$, the cardinality
$D_t=|\CT^n_t|$ is lower bounded as \cite{CT91}
\begin{equation}
\label{cc4} D_t \geq 2^{n [H(p) - \eta(\delta)]}
\end{equation}
and the function $\eta(\delta) \rightarrow 0$ as $\delta \rightarrow
0$.

The above concepts generalize to the quantum setting by virtue of
the spectral theorem. Let $\rho=\sum_{x\in\CX}p_x \ket{x}\bra{x}$ be
the spectral decomposition of a given  density matrix $\rho$. In
other words, $\ket{x}$ is the eigenstate of $\rho$ corresponding to
eigenvalue $p_x$. The von Neumann entropy of the density matrix
$\rho$ is
$$H(\rho)=-\tr{\rho\log\rho}=H(p).$$
Define the type projector
$$\Pi^n_t=\sum_{x^n\in\CT^n_{t}}\ket{x^n}\bra{x^n}.$$ The density
operator proportional to the type projector is
$\pi_t={D_t}^{-1}{\Pi^n_t}$. The typical subspace associated with
the density matrix $\rho$ is defined as
$$
\FT=\sum_{x^n\in\CT_{p,\delta}^n}\ket{x^n}\bra{x^n}=\sum_{t\in\tau_{\delta}}\Pi^n_t.
$$
Properties analogous to (\ref{cc1}) -- (\ref{cc4}) hold \cite{NC00}.
For any $\ep,\delta>0$, and all sufficiently large $n$ for which
\beq \tr{\rho^\on \FT}  \geq  1-\ep  \label{q1} \eeq \beq 2^{-n\lb
H(\rho)+c\delta\rb}\FT  \leq  \FT \rho^\on \FT  \leq 2^{-n\lb
H(\rho)-c\delta \rb}\FT \label{q2}, \eeq \beq \tr{\FT} \leq 2^{n\lb
H(\rho)+c\delta\rb}  \label{q3} \eeq for some constant $c$. For
$t\in \tau_\delta$ and for sufficiently large $n$, the dimension of
the type projector $\Pi^n_t$ is lower bounded as \beq \label{cc7}
\tr\Pi^n_t \geq 2^{n [H(\rho) - \eta(\delta)]} \eeq and the function
$\eta(\delta) \rightarrow 0$ as $\delta \rightarrow 0$.

For a multipartite state $\rho^{ABC}$, we write $H(A)_\rho =
H(\rho^A)$, etc. We omit the subscript if the state is clear from
the context. Define the {quantum mutual information} by
$$I(A;B)=H(A)+H(B)-H(AB)$$
and the {quantum conditional mutual information} by
$$I(A;C|B) = H(AB)+H(BC)-H(ABC)-H(B).$$
These are non-negative by strong subadditivity~\cite{LR73}. If
$I(A;B) = 0$ then
$$
I(A;C|B) = I(A;CB)
$$
is easy to verify.

The set of generalized Pauli matrices $\{U_m\}_{m \in [d^2]}$ is
defined by $U_{l \cdot d+k} = \hat{Z}_{d}(l)\hat{X}_{d}(k)$ for
$k,l=0,1,\cdots,d-1$ and \beq
\begin{split}
\label{gpm}
\hat{X}_{d}(k)&=\sum_{s}\ket{s}\bra{s+k}=\hat{X}_{d}(1)^k,  \\
\hat{Z}_{d}(l)&=\sum_{s}e^{i2\pi
sl/d}\ket{s}\bra{s}=\hat{Z}_{d}(1)^l.
\end{split}
\eeq The $+$ sign denotes addition modulo $d$.

We will always use $\ket{\Phi}$ to represent the maximally entangled
state. Then the maximally entangled state $\ket{\Phi}^{AB}$ on a
pair of $d$-dimensional quantum systems $A$ and $B$ is given as:
\begin{equation}
\ket{\Phi}^{AB}= \frac{1}{\sqrt{d}}\sum_{i=1}^{d}\ket{i}^A\ket{i}^B.
\label{maxen}
\end{equation}
We have the following result (see \cite{BSST01} for a proof):
\begin{equation}
\label{randomo}
\frac{1}{d^2}\sum_{m=1}^{d^2}(U_m \otimes I)\Phi^{AB}(U_m^{\dagger}\otimes I) \\
=\pi^A \otimes \pi^B,
\end{equation}
where $\pi^A =\pi^B= \frac{I}{d}$. We will also need the following
equality: \beq (I \otimes U) \ket{\Phi} = (U^{tr} \otimes I)
\ket{\Phi} \label{car} \eeq for any operator $U$, and $U^{tr}$
denotes transposition of $U$.

Next is a coherent version of the gentle operator lemma
(\cite{Winter99}, Lemma 9). It states that a measurement which is
likely to be successful in identifying a state tends not to
significantly disturb the state.

\begin{lemma}[Gentle coherent measurement] 
\label{cor1} Let $\{\rho_k^A\}_{k\in[K]}$ be a collection of density
operators and $\{\Lambda_k\}_{k\in[K]}$ be a set of POVMs on quantum
system $A$ such that
$$ \tr{\rho_k \Lambda_k} \geq 1-\ep $$
for all $k$. Let $\ket{\phi_k}^{RA}$ be a purification of
$\rho^A_k$. Then there exists an isometric quantum operation $\CD:A
\rightarrow A J$ such that
$$
\|(I^R \otimes \CD)(\phi_k^{RA}) - \phi_k^{RA} \otimes \proj{k}^J \|
\leq \sqrt{8\ep}.
$$
\end{lemma}
\begin{IEEEproof}
Every POVM can be written as an isometry followed by projective
measurement on a subsystem. In particular, there exists an isometry
$\CD:A \rightarrow A J$ such that
$$
(I^R\otimes\CD) \ket{\phi}^{RA} = \sum_j [(I^R \otimes
\sqrt{\Lambda_j}) \ket{\phi}^{RA}] \ket{j}^J.
$$
Thus
\begin{equation}
\begin{split}
\bra{k} \bra{\phi_k} (I \otimes \CD) \ket{\phi_k}
& = \bra{\phi_k} ( I \otimes \sqrt{\Lambda_k})  \ket{\phi_k} \\
& \geq   \bra{\phi_k} (I \otimes {\Lambda_k})  \ket{\phi_k} \\
& = \tr{\rho_k \Lambda_k} \\
& \geq 1-\ep.
\end{split}
\end{equation}
The first inequality uses that $\Lambda_k \leq \sqrt{\Lambda_k}$
when $0 \leq \Lambda_k \leq I$. The statement of the lemma follows
from the fact that for pure states $\ket{\zeta}$ and $\ket{\psi}$,
$$\| \zeta - \psi \| = 2 \sqrt{1 - |\langle{\zeta}\ket{\psi} |^2}.$$
\end{IEEEproof}

The packing lemma below will prove to be a powerful tool in quantum
information theory. The technique used here is simple, directly
analogous to the classical coding theorem.
\begin{lemma}[Packing]
\label{packing} We are given an ensemble $\{\lambda_m,
\sigma_m\}_{m\in\CS}$ with average density operator $$\sigma =
\sum_{m \in \CS} \lambda_m \sigma_m.$$ Assume the existence of
projectors $\Pi$ and $\{\Pi_m\}_{m \in \CS}$  with the following
properties:
\begin{eqnarray}
\tr \sigma_m \Pi_m & \geq & 1 - \epsilon,\label{c2} \\
\tr \sigma_m \Pi & \geq & 1 - \epsilon, \label{c1} \\
\tr \Pi_m &  \leq  & d, \label{c3} \\
\Pi \sigma \Pi  &  \leq  & D^{-1} \Pi \label{c4}
\end{eqnarray}
for all $m \in \CS$ and some positive integers $D$ and $d$. Let $N =
\lfloor \gamma D/d \rfloor$ for some $0 < \gamma < 1$ where $\lfloor
r \rfloor$ represents the largest integer less than $r$. Then there
exists a map $f:[N]\to\CS$, and a corresponding set of POVMs
$\{\La_k\}_{k \in [N]}$ which reliably distinguishes between the
states $\{\sigma_{f(k)}\}_{k\in[N]}$ in the sense that
$$\tr\sigma_{f(k)} \La_k \geq 1-4(\ep+\sqrt{8\ep})-8\gamma$$
for all $k\in[N]$.
\end{lemma}
\begin{IEEEproof} See Appendix \ref{Appendix:packing}.
\end{IEEEproof}
\begin{lemma}
\label{lem5} If $\ket{\psi}^{ABE}$ is a pure state then
$$
H(B|E)_\psi = - H(B|A)_\psi.
$$
\end{lemma}
\begin{IEEEproof} Since $\ket{\psi}^{ABE}$ is pure, we have
$H(A)_\psi=H(BE)_\psi$ and $H(E)_\psi=H(AB)_\psi$. Then \beq
\begin{split}
H(B|E)_\psi &= H(BE)_\psi-H(E)_\psi \\
&=H(A)_\psi-H(AB)_\psi \\
&= - H(B|A)_\psi.
\end{split}
\eeq
\end{IEEEproof}

\begin{lemma}
\label{lem6} For any state $\sigma^{ABE}$,
$$
I(A;B)_\sigma \leq  H(B)_\sigma +  H(B|E)_\sigma.
$$
\end{lemma}
\begin{IEEEproof} Introduce a reference system $R$ that purifies the state
$\sigma^{ABE}$, then \begin{equation}
\begin{split}
I(A;B)_\sigma&=H(B)_\sigma-H(B|A)_\sigma  \\
&= H(B)_\sigma+H(B|ER)_\sigma  \\
&\leq  H(B)_\sigma +  H(B|E)_\sigma.
\end{split}
\end{equation}
The first equality follows from the definition of quantum mutual
information. The second equality follows from Lemma \ref{lem5}. The
first inequality uses the fact that conditioning reduces entropy
\cite{LR73}.
\end{IEEEproof}

\section{Main Result}
\label{III}
\subsection{Two party entanglement-assisted coding}
Before attacking the multiuser problem we give a new proof of the
two-party entanglement-assisted direct coding theorem. This theorem
was first proved in \cite{BSST01} and subsequently in \cite{Hol01a}.
Both proofs invoke the HSW theorem.
The HSW theorem uses the method of conditionally typical subspaces.
We give a direct proof based on the packing lemma which only uses
typical subspaces. The proof perhaps sheds more light on why
achievable rates take on the form of mutual information.
Furthermore, our proof provides new properties ( ii) and iii) below)
that serve as a bridge to the proof of multiparty coding theorem.

\begin{figure}
\centering
\includegraphics[width=0.4\textwidth]{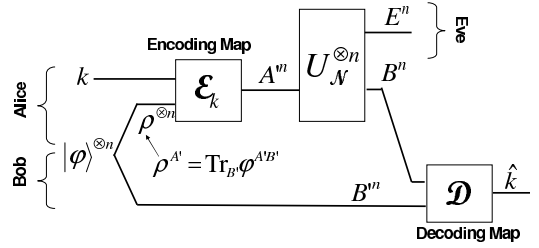}
\caption{Two-party entanglement-assisted communication}
\label{fig:two}
\end{figure}
As shown in Fig. \ref{fig:two}, Alice and Bob are connected by a
large number $n$ uses of the quantum channel $\CN: A' \rightarrow
B$. Alice controls the channel input system $A'$ and Bob has access
to the channel output $B$. They also have entanglement in the form
of $n$ copies of some pure bipartite state $\varphi^{A'B'}$. Any
such state is determined upto a local unitary transformation by the
local density operator $\rho^{A'} = \tr_{B'} \varphi^{A'B'}$. Alice
and Bob use these resources to communicate, in analogy to superdense
coding \cite{BW92}. Based on her message Alice performs a quantum
operation on her share of the entanglement. She then sends it
through the quantum channel. Bob performs a decoding measurement on
the channel output plus his share of the entanglement. They endeavor
to maximize the communication rate.

We formalize the above information processing task. Define an
$[n,R,\rho,\ep]$ entanglement-assisted code by
\begin{itemize}
\item
a set of unitary encoding maps $\{\CE_k\}_{k \in [2^{nR}]}$ acting
on ${A'}^{n}:={A'_1\dots A'_n}$ for Alice;
\item
Bob's decoding instrument ${\bf D} = \{\CD_k\}_{k \in [2^{nR}]}$
acting on $B^n B'^n$.
\end{itemize}
such that for all $k \in  [2^{nR}]$
\begin{enumerate}[i)]
\item
$ \tr\{ [\CD_k \circ((\CN^{\otimes n} \circ \CE_k) \otimes I)]
(\varphi^{\otimes n})\} \geq 1 - \epsilon; $
\item the encoded density operator satisfies
$\CE_k(\rho^{\otimes n}) = \rho^{\otimes n}$;
\item
\begin{multline*} \left\|  [(\CD \otimes I^{E^n}) \circ(({U_\CN}^{\otimes n} \circ
\CE_k) \otimes I) - ({U_\CN}^{\otimes n}\otimes I)]
(\varphi^{\otimes n}) \right\| \\ \leq \epsilon\end{multline*} where $\circ$
represents composition of two maps.
\end{enumerate}

Condition i) means that Bob correctly decodes Alice's message with
high probability. This condition suffices for two-party
entanglement-assisted communication. The remaining two properties,
which were not shown in \cite{BSST01,Hol01a}, are important for the
multiparty generalization. Condition ii) means that Alice always
inputs a tensor power state into the channel. Condition iii) says
that the encoding and decoding operations in effect cancel each
other out. So it is as if Alice just sent the state $\rho^{\otimes
n}$ down the channel without any coding. In reality, she has also
managed to convey the message to Bob.

\begin{theorem}
\label{thm1} Define $\theta^{AB} = (I \otimes \CN) \varphi^{A A'}$
and $R = I(A;B)_\theta$. For every $\ep,\delta > 0$ and n
sufficiently large, there exists an $[n,R-\delta,\rho,\ep]$
entanglement-assisted code.
\end{theorem}

\begin{IEEEproof}
Let $t(1), \dots, t(a)$ be an ordering of the distinct types
$t^{x^n}$. Define the maximally mixed state $\pi^n_\alpha =
1/d_\alpha \Pi^n_{t(\alpha)}$, where $d_\alpha = \tr
\Pi^n_{t(\alpha)}$. Define $\ket{\Phi_\alpha}$ to be the maximally
entangled state on a pair of $d_\alpha$-dimensional quantum systems
$A'^n$ and $B'^n$
\begin{equation}
\ket{\Phi_\alpha}^{{A'}^n {B'}^n} = \frac{1}{\sqrt{d_\alpha}}
\sum_{x^n\in\CT^n_{t(\alpha)}} \ket{x^n}^{{A'}^n}\ket{x^n}^{{B'}^n}.
\end{equation}
In the beginning Alice and Bob share the entangled state \beq
\begin{split}
\ket{\Psi}^{{A'}^n {B'}^n} &=  \ket{ \varphi}^{\otimes n}\\
&=\sum_{\alpha}\sqrt{p_\alpha}\ket{\Phi_\alpha},
\end{split}
\eeq where $p_\alpha = \sum_{x^n \in \CT^n_{t(\alpha)}} p^n(x^n)$.
The type projectors $\Pi^n_{t(\alpha)}$ induce a decomposition of
the Hilbert space $\CH^{\otimes n}$ of ${A'}^n$ (correspondingly of
$ {B'}^n$) into a direct sum
$$
\CH^{\otimes n} = \bigoplus_{\alpha = 1}^{a} \CH_{t(\alpha)}.
$$
Let $\CG=\{(g_1,g_2,\cdots,g_a):g_\alpha \in [d_\alpha^2
],\alpha\in[a] \}$, $\CB=\{(b_1,b_2,\cdots,b_a):b_\alpha \in \{0,1
\}\}$, and $\CS=\CG\times\CB$. Every element $s^a\in\CS$ is uniquely
determined by $g^a\in\CG$ and $b^a\in\CB$. Given an element
$s^a\in\CS$, define a unitary operation $\U$ to be
\begin{equation}
\U\equiv U_{g^a,b^a}=\bigoplus_{\alpha = 1}^{a} (-1)^{b_\alpha}
U_{g_\alpha} \label{ucons}
\end{equation}
where $\{U_{g_\alpha}\}$ are the $d_\alpha^2$ generalized Pauli
operators (\ref{gpm}) defined on $\CH_{t(\alpha)}$. Define \beq
\label{sigthe}
\begin{split}
\S^{B^n {B'}^n}&:=(\CNn\otimes I)\lb(\U\otimes I)\Psi^{{A'}^n{B'}^n}(\U^{\dagger} \otimes I)\rb  \\
&=  (I \otimes \U^{tr})  \theta^{\otimes n} (I \otimes \U^{*}).
\end{split}
\eeq The last equality follows from (\ref{car}). Let $\sigma$ to be
the average of $\S$ over $\CS$, then we get (\ref{sigma_ave}). 
\begin{figure*}
\beq
\begin{split}\label{sigma_ave}
\sigma&=\frac{1}{|\CS|}\sum_{s^a\in\CS}\S \\
&=\frac{1}{|\CB||\CG|}\sum_{g^a\in\CG}\sum_{b^a\in\CB}
\sum_{\alpha,\alpha'}\sqrt{p_\alpha p_{\alpha'}} (\CNn\otimes
I)\lb(U_{g^a,b^a}\otimes I)\ket{\Phi_\alpha}
\bra{\Phi_{\alpha'}}(U_{g^a,b^a}^{\dagger} \otimes I)\rb. \\
&=\sum_{\alpha}p_\alpha \BL \CNn (\pit)\otimes\pit\BR.
\end{split}
\eeq 
\end{figure*}
The last equality comes from (\ref{t=t}) and (\ref{tneqt})
below. When $\alpha=\alpha'$, 
\begin{align}
\label{t=t}
&\frac{1}{|\CB||\CG|}\sum_{g^a\in\CG}\sum_{b^a\in\CB}p_\alpha
(\CNn\otimes I)\lb(U_{g^a,b^a}\otimes I)\Phi_\alpha(U_{g^a,b^a}^{\dagger} \otimes I)\rb \nonumber \\
&=(\CNn\otimes I)
 \frac{1}{|\CG|}\sum_{g_1}\cdots\sum_{g_a}p_\alpha(U_{g_\alpha} \otimes I)\Phi_\alpha
( U_{g_\alpha}^{\dagger}\otimes I) \nonumber \\
&=(\CNn\otimes I) p_\alpha (\pit\otimes\pit).
\end{align} 
The last equality follows from (\ref{randomo}). When $\alpha
\neq \alpha'$, we get (\ref{tneqt}).
\begin{figure*}[t]
\beq
\begin{split}
\label{tneqt}
&\frac{1}{|\CB||\CG|}\sum_{g^a\in\CG}\sum_{b^a\in\CB}\sqrt{p_\alpha
p_{\alpha'}} (\CNn\otimes I) \lb(U_{g^a,b^a}\otimes
I)\ket{\Phi_\alpha}\bra{\Phi_{\alpha'}}(U_{g^a,b^a}
^{\dagger} \otimes I)\rb \\
&= \frac{1}{d_\alpha^2 d_{\alpha'}^2}\sqrt{p_\alpha p_{\alpha'}}
\sum_{b_\alpha
b_{\alpha'}}\frac{(-1)^{b_\alpha+b_{\alpha'}}}{4}\left\{
\sum_{g_\alpha g_{\alpha'}} (\CNn\otimes I)\lb(U_{g_\alpha}\otimes
I)\ket{\Phi_\alpha}\bra{\Phi_{\alpha'}}
(U_{g_{\alpha'}}^{\dagger} \otimes I)\rb \right\}  \\
&=0.
\end{split}
\eeq 
\end{figure*}
Define the projectors on ${B'}^n B^n$ \beq \label{proje}
\begin{split}
\P&=(I\otimes \U^{tr})\, \Pi_{\theta,\delta}^{n} \, (I\otimes \U^{*}), \\
\Pi&=\Pi_{\CN(\rho),\delta}^{n}\otimes\FT.
\end{split}
\eeq The following properties are proved in Appendix
\ref{appendix1}. For all $\epsilon > 0, \delta > 0$ and all
sufficiently large $n$,
\begin{eqnarray}
\tr{\S \Pi}&\geq& 1-\epsilon  \label{pl} \\
\tr{\S \P}&\geq& 1-\epsilon \label{p2}\\
\tr{\P} &\leq& 2^{n\lb H(AB)_\theta+c\delta \rb}  \label{p3}\\
\Pi\sigma\Pi &\leq& 2^{n\lb H(A)_\theta + H(B)_\theta +c\delta
\rb}\Pi.  \label{p4}
\end{eqnarray}
Let $\lambda_{s^a}=\frac{1}{|\CS|}$ and 
 $R =  I(A;B)_\theta - (2c + 1) \delta$.
We now apply the packing lemma to the ensemble
$\{\lambda_{s^a},\S\}_{s^a\in\CS}$ and projectors $\Pi$ and $\P$.
Thus there exist a map $f: [2^{nR}] \to \CS$ and a POVM
$\{\Lambda_{k}\}_{k\in[2^{nR}]}$ such that
\begin{equation}
\label{inF} \tr{\sigma_{f(k)}\Lambda_{k}} \geq 1-\epsilon',
\end{equation}
with
$$
\epsilon' = 4(\epsilon + \sqrt{8\epsilon}) + 16 \times 2^{-n
\delta}.
$$
Define the encoding operation by  $\CE_{k} = U_{f(k)}$. Including
the environment system, the state of $B^n{B'}^nE^n$ after the
application of the channel $U_\CN$ is \beq
\begin{split}
\ket{\Upsilon_k}^{B^n{B'}^nE^n} &=({U_\CN}^{\otimes n} \otimes I)
(U_{f(k)}\otimes I)\ket{\Psi}^{{A'}^n{B'}^n}\\
& = ({U_\CN}^{\otimes n} \otimes U^{tr}_{f(k)})
\ket{\Psi}^{{A'}^n{B'}^n}.
\end{split}
\eeq $\ket{\Upsilon_k}$ is a purification of $\sigma_{f(k)}$. By
Lemma \ref{cor1}, there exists an isometry $\CD': B^n B'^n
\rightarrow B^n B'^n J$ such that
$$
\| (I \otimes \CD')(\Upsilon_k) - \Upsilon_k \otimes \proj{k}^J \|
\leq \sqrt{8\ep'}.
$$
Bob performs the controlled unitary
$$
W^{J {B'}^n}  = \sum_k \proj{k}^J \otimes (U^{*}_{f(k)})^{{B'}^n}.
$$
Defining $\CD'' = (W \otimes I^{B^n}) \circ \CD'$, this implies
\begin{equation}
\label{cohir} \left\| (I \otimes \CD'')(\Upsilon_k) -
[(U_\CN^{\otimes n} \otimes I) (\varphi^{\otimes n})] \otimes
\proj{k} \right\| \leq \sqrt{8\ep'}.
\end{equation}
The instrument ${\bf D}=\{\CD_k\}$ is defined by $\CD''$ followed by
a von Neumann measurement of the system $J$. Equation (\ref{cohir})
expresses the fact that the classical communication being performed
is almost decoupled from  all the quantum systems involved in the
protocol, including ancillas and the inaccessible environmnent. We
remark that this guarantees the ability to ``coherify'' the protocol
in the sense of \cite{DHW03}.

Condition i) in the form
$$
\tr\left\{ \left[\CD_k \circ((\CN^{\otimes n} \circ \CE_k) \otimes
I)\right] (\varphi^{\otimes n})\right\} \geq
 1 - \ep'
$$
is immediate from (\ref{inF}). Condition ii) follows from the
construction (\ref{ucons}). Condition iii)  in the form
$$
\left\|  [(\CD \otimes I^{E^n}) \circ((U_\CN^{\otimes n} \circ
\CE_k) \otimes I) - (U_\CN^{\otimes n}  \otimes I)]
(\varphi^{\otimes n}) \right\| \leq \sqrt{8 \epsilon'}
$$
follows from (\ref{cohir}).
\end{IEEEproof}

\subsection{Remark on the HSW theorem}
\label{ska} Suppose that Alice and Bob are connected by a special
\emph{cq channel} of the form
$$
\CN = \CN' \circ \Delta,
$$
where $\Delta$ is the dephasing channel
$$
\Delta:\rho \rightarrow \sum_x \proj{x} \rho  \proj{x}.
$$
A $\{c \to q \}$ channel is equivalent to one with classical inputs
and quantum outputs. The HSW coding theorem states that rates $ R =
I(A;B)_\theta, $ $\theta^{AB} = (I \otimes \CN) \varphi^{A A'}$ are
achievable even without entanglement assistance. We show that this
fact follows from our construction in two steps.

The first step is to replace the entanglement used by classical
common randomness. Observe that the encoding operations $\U$ all
satisfy
$$
\Delta^{\otimes n} \circ \U = \Delta^{\otimes n} \circ \U  \circ
\Delta^{\otimes n}.
$$
This follows from the corresponding property of the generalized
Pauli operators (\ref{gpm}). Hence for cq channels $\CN$ \beq
\begin{split}
\sigma_{f(k)}  & =  [(\CN^{\otimes n} \circ \CE_k) \otimes I] (\varphi^{\otimes n}) \\
& =   [(\CN^{\otimes n} \circ \CE_k \circ \Delta^{\otimes n}) \otimes I] (\varphi^{\otimes n}) \\
& = [ (\CN^{\otimes n} \circ \CE_k) \otimes I]
(\bar{\varphi}^{\otimes n}),
\end{split}
\eeq where
$$
\bar{\varphi}  =  (\Delta  \otimes I) \varphi = \sum_x p_x \proj{x}
\otimes \proj{x}
$$
is the dephased version of $\varphi$. The state
$\bar{\varphi}^{\otimes n}$ can be constructed from classical common
randomness like that used in Shannon's original coding theorem.

The second step is showing that common randomness is not needed. The
argument parallels the derandomization step from the proof of the
packing lemma (Appendix \ref{Appendix:packing}). We have thus
recovered the HSW coding theorem.

The benefit of the above proof is its close analogy to Shannon's
joint typicality decoding. We only made use of typical subspaces and
not conditionally typical subspaces.

\subsection{Multiple-Access Channel}
We turn to the communication scenario with two senders, Alice and
Bob, and one receiver, Charlie. They are connected by a large number
$n$ of uses of the \emph{multiple-access} quantum channel
$\CM:A'B'\to C$. Alice and Bob control the channel input systems
$A'$ and $B'$, respectively. Charlie  has access to the channel
output $C$. Each sender also shares unlimited entanglement with the
receiver, in the form of arbitrary pure states
$\ket{\Gamma_1}^{AC_A}$ and $\ket{\Gamma_2}^{BC_B}$. The system $A$
is held by Alice, $B$ by Bob, and $C_AC_B$ by Charlie. Based on her
message Alice performs a quantum operation on her share of the
entanglement, and likewise for Bob. These are then sent through the
quantum channel. Charlie performs a decoding measurement on the
channel output plus his share of the entanglement. Now both Alice's
and Bob's communication rates need to be optimized.

We formalize the above information processing task. Define an
$(n,R_1,R_2, \ep)$ entanglement-assisted code by
\begin{itemize}
\item two sets of encoding cptp maps:
$\{\CE^1_k\}_{k\in [2^{nR_1}]}$ taking $A$ to $A'^n$ for Alice, and
$\{\CE^2_l\}_{l\in [2^{nR_2}]}$ taking $B$ to $B'^n$ for Bob ;
\item Charlie's decoding POVM
$\{\Lambda_{k,l}\}_{k\in [2^{nR_1}],l\in [2^{nR_2}]}$ on $C_AC_BC$,
\end{itemize}
such that
\begin{multline} \label{macc} \tr\{\Lambda_{k,l}
[((\CM^\on\circ(\CE^1_k\otimes\CE^2_l))\otimes I^{C_AC_B})
(\Gamma_1^{AC_A} \otimes  \Gamma_2^{BC_B})]\}\\  \geq 1-\ep.
\end{multline}

We say that $(R_1,R_2)$ is an achievable rate pair if for all
$\epsilon>0, \delta>0$ and sufficiently large $n$ there exists an
$(n,R_1-\delta,R_2-\delta,\ep)$ entanglement-assisted code. The
entanglement-assisted {\it capacity region} $C_{E}(\CM)$ is defined
to be the closure of the set of all achievable rate pairs.

\begin{theorem}
\label{t2} Consider a quantum multiple access channel $\CM:A'B'\to
C$. For some states $\rho_{1}^{A'}$ and $\rho_{2}^{B'}$ define \beq
\theta^{A B C}= (I^{AB} \otimes \CM) (\varphi_{1}^{A
A'}\otimes\varphi_{2}^{B B'}) \label{theeta} \eeq where
$\ket{\varphi_{1}}^{A A'}$ and $\ket{\varphi_{2}}^{B B'}$ are
purifications of $\rho_{1}^{A'}$ and $\rho_{2}^{B'}$ respectively.
Define the two-dimensional region $C_E(\CM,\rho_1, \rho_2)$, shown
in Fig.~\ref{fig:region}, by the set of pairs of nonnegative rates
$(R_1,R_2)$ satisfying \beq
\begin{split}
\label{cr}
R_1 & \leq I(A; C|B)_{\theta}  \\
R_2 & \leq I(B; C|A )_{\theta} \\
R_1+R_2 & \leq I(AB; C)_{\theta}.
\end{split}
\eeq Define $\tilde{C}_E(\CM)$ as the union of the $C_E(\CM,\rho_1,
\rho_2)$ regions taken over all
 states ${\rho_1, \rho_2}$. Then the
entanglement-assisted capacity region $C_E(\CM)$ is given by the
regularized expression
\begin{equation}
C_E(\CM) = \overline{ \bigcup_{n = 1}^{\infty}\frac{1}{n}
\tilde{C}_E(\CM^{\otimes n}) }
\end{equation}
where the bar indicates taking closure. There is an additional
single-letter upper bound on the sum rate \beq \label{slub} R_1 +
R_2 \leq \max_{\rho_1, \rho_2} I(AB; C)_{\theta}. \eeq

\begin{figure}
\centering
\includegraphics[width=0.4\textwidth]{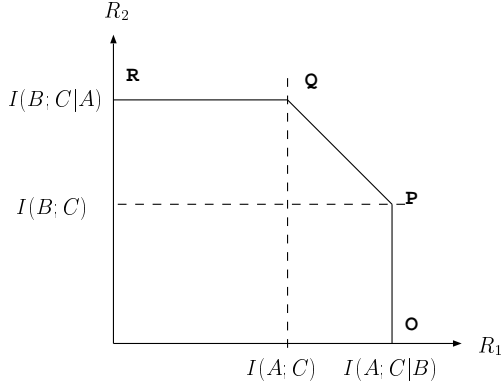}
\caption{Capacity region of multiple access channel for fixed input
states $\rho_1$ and $\rho_2$} \label{fig:region}
\end{figure}
\end{theorem}

\begin{IEEEproof}(direct coding theorem)
Let the entanglement be given in a tensor power form, as in Theorem
1. Define a  $[n,R_1,R_2,\rho_1, \rho_2,\ep]$  entanglement-assisted
code as a special case of an $(n,R_1,R_2,\ep)$ code: specify
$\Gamma_1 = \phi_1^\on$ and $\Gamma_2 = \phi_2^\on$, and identify $A
:= A'^n$ and $B := B'^n$.

To show the achievability of every rate pair $(R_1,R_2)$ in the
convex hull of the $C_E(\CM, \rho_1, \rho_2)$, it suffices to show
that the corner points are achievable.
Once we show that, the non-corner points can be achieved by
time-sharing (see, e.g., \cite{DS03}). Consider the corner point Q.
For all $\ep>0, \delta>0$ and $n$ sufficiently large, we show below
that there exists a $[n,I(A;C)_\theta -\delta, I(B;C|A)_\theta
-\delta,\rho_1,\rho_2,\ep]$ entanglement-assisted code
$(\CE^{1},\CE^2,\CD)$.

The point Q corresponds to the maximum rate that at which Alice can
send as long as Bob sends at his maximum rate. This is the rate that
is achieved when Bob's input is considered as noise for the channel
from Alice to Charlie. From the two party direct coding theorem,
Alice can send at a rate $I(A;C)$ and Charlie can decode the message
with arbitrarily low probability. Charlie then knows which encoding
operation Alice used and can subtract its effect from the channel.
Therefore, Bob can achieve the rate $I(B;C|A)$. This outlines the
proof of the achievability of point Q.

Define the channel $\CN_1:A'\rightarrow C$ by
$$ \CN_1:\omega \mapsto\CM(\omega \otimes\rho_2).$$
$\CN_1^{\on}$ is the effective channel from Alice to Charlie when
Bob's input to $\CM^{\on}$ is $\rho_{2}^\on$. Define
$\hat{\CN}_1:A'\rightarrow C_B C$ by
$$ \hat{\CN}_1:\omega \mapsto (I \otimes \CM) (\omega \otimes \varphi_2). $$
Observe that $\hat{\CN}_1$ is an extension of $\CN_1$. Hence it is a
restriction of  $U_{\CN_1}$.

Define the channel $\CN_2:B'\rightarrow C_A C$ by
$$ \CN_2:\omega \mapsto (I \otimes \CM) (\varphi_{1}\otimes \omega). $$
$\CN_2^{\on}$ is effective the channel from Bob to Charlie if Alice
simply inputs the $A'$ part of the entangled state/purification
$(\varphi_1^{A'C_A})^\on$ without encoding.

Fix $\ep>0, \delta>0$. Define $R_1 = I(A;C)_\theta -\delta$ and $R_2
=I(B;C|A)_\theta -\delta$, with $\theta$ defined in (\ref{theeta}).
By Theorem 1, for sufficiently large $n$ there exists an
$[n,R_1,\rho_1,\ep]$ entanglement-assisted code $(\CE^1,\CD^1)$ for
$\CN_1$ and an $[n,R_2,\rho_2,\ep]$ entanglement-assisted code
$(\CE^2,\CD^2)$ for $\CN_2$ such that for all $k \in [2^{nR_1}], l
\in [2^{nR_2}]$,
\begin{enumerate}[i)]
\item  \label{nma1}
$ \tr\{ [\CD^1_k \circ((\CN_1^{\otimes n} \circ \CE^1_k) \otimes
I^{C_A})] (\varphi_1^{\otimes n})\} \geq 1 - \epsilon; $
\item  \label{nma2}
\begin{multline*} \left\| [(\CD^1 \otimes I) \circ((\hat{\CN}_1^{\otimes n} \circ
\CE^1_k) \otimes I^{C_A}) - (\hat{\CN}_1^{\otimes n}  \otimes
I^{C_A})] (\varphi_1^{\otimes n}) \right\|\\ \leq \epsilon; \end{multline*}
\item  \label{nmb1}
$ \tr\{ [\CD^2_l \circ((\CN_2^{\otimes n} \circ \CE^2_l) \otimes
I^{C_B})] (\varphi_2^{\otimes n})\} \geq 1 - \epsilon; $
\item   \label{nmb2}
the encoded density operator satisfies $\CE^2_l(\rho_2^{\otimes n})
= \rho_2^{\otimes n}$.
\end{enumerate}
We now define our code for the multiple access channel $\CM$. Alice
and Bob encode according to $\{\CE^1_k\}$ and $\{\CE^2_l\}$,
respectively. Define the instrument ${\bf D}=\{\CD_{k,l}\}$ on
$CC_AC_B$ by
$$
\CD_{k,l} = \CD_l^2  \circ (\CD_k^1 \otimes I^{C_B}).
$$
Then Charlie's decoding POVM $\{\Lambda_{k,l}\}$ is the restriction
of $\{\CD_{k,l}\}$. Examining the success probability of decoding
Alice's message $k$: \beq \label{macc2}
\begin{split}
& \tr\{(\CD_k^1 \otimes I^{C_B}) \circ
((\CM^\on\circ(\CE^1_k\otimes\CE^2_l))\otimes I^{C_AC_B})
(\varphi_1^{\otimes n} \otimes \varphi_2^{\otimes n}) \} \\
& = \tr\{\CD_k^1  \circ ((\CM^\on\circ(\CE^1_k\otimes\CE^2_l))
\otimes I^{C_A})(\varphi_1^{\otimes n} \otimes \rho_2^{\otimes n}) \} \\
& =\tr\{\CD_k^1  \circ ((\CM^\on\circ(\CE^1_k\otimes
I^{B'^n}))\otimes I^{C_A})( \varphi_1^{\otimes n} \otimes 
\rho_2^{\otimes n}) \} \\
& =\tr\{ [\CD^1_k \circ((\CN_1^{\otimes n} \circ \CE^1_k) \otimes I)](\varphi_1^{\otimes n})\} \\
& \geq  1-\ep.
\end{split}
\eeq The second equality follows from  \ref{nmb2}) and the third
from  \ref{nma1}).

Next examining the success probability of decoding Bob's message
$l$: Rewrite \ref{nma2}) in terms of $\CM$:
\begin{multline*}
\| [(\CD^1  \otimes I^{C_B})\circ((\CM^{\otimes n} \circ
(\CE^1_k  \otimes  I^{B'^n} )) \otimes I^{C_AC_B})\\ - (\CM^{\otimes
n}\otimes I^{C_AC_B }) ] (\varphi_1^{\otimes n} \otimes
\varphi_2^{\otimes n}) \| \leq \epsilon;
\end{multline*}
Since $\CE^2_l$ is unitary and satisfies \ref{nmb2}),
\begin{multline*}
\| [(\CD^1  \otimes I^{C_B})\circ((\CM^{\otimes n} \circ
(\CE^1_k  \otimes  \CE^2_l ) ) \otimes I^{C_AC_B})\\- ((\CM^{\otimes
n} \circ ( I^{A'^n}  \otimes  \CE^2_l ))   \otimes I^{C_AC_B }) ]
(\varphi_1^{\otimes n} \otimes \varphi_2^{\otimes n}) \| \leq
\epsilon;
\end{multline*}
Rewrite \ref{nmb1}) in terms of $\CM$
\begin{multline*}
\tr\{ [\CD^2_l   \circ((\CM^{\otimes n} \circ ( I^{A'^n} \otimes
\CE^2_l) ) \otimes I^{C_AC_B })] (\varphi_1^{\otimes n} \otimes
\varphi_2^{\otimes n} ) \} \\
 \geq 1 - \epsilon.
\end{multline*}
Define
\begin{multline*}
\Omega^{CC_AC_B}  \\ = (\CD^1 \otimes I^{C_B}) \circ((\CM^\on\circ
(\CE^1_k\otimes\CE^2_l))\otimes I^{C_AC_B})( \varphi_1^{\otimes n}
\otimes \varphi_2^{\otimes n})
\end{multline*}
Hence
$$
\tr [\CD^2_l \, \Omega^{CC_AC_B}] \geq 1 - 2 \epsilon.
$$
Now (\ref{macc}) follows. This concludes the achievability of point
Q.

Corner point P can be shown in the same manner. Corner point R
corresponds to the maximum rate achievable from Bob to Charlie when
Alice is not sending any information. The proof is obvious since we
can assume that Alice is throwing the same state into the channel
all the time. The corner point O follows from the same reasoning.
This concludes the proof of direct coding theorem.
\end{IEEEproof}

\medskip

{\bf Remark}. The entanglement assistance may be phrased in terms of
tensor powers of ebit states $\ket{\Phi_+} = \frac{1}{\sqrt{2}}(
\ket{0} \ket{0} +  \ket{1} \ket{1})$ instead of the arbitrary
$\ket{\Gamma_1}$ and $\ket{\Gamma_2}$. The protocol achieving the
corner points of the region $C_E(\CM, \rho_1, \rho_2)$ uses
$\ket{\Gamma_1} = \ket{\phi_1}^{\otimes n}$  and $\ket{\Gamma_2} =
\ket{\phi_2}^{\otimes n}$. By entanglement dilution \cite{HL02},
$\ket{\Gamma_1}$ may be asymptotically obtained from an ebit rate of
$E_1 = H(A)_\theta$ shared between Alice and Charlie.  Likewise
$\ket{\Gamma_2}$ may be asymptotically obtained from an ebit rate of
$E_2 = H(B)_\theta$ shared between Bob and Charlie. Entanglement
dilution additionally requires an arbitrarily small rate of
classical communication. This resource is obtained by applying the
HSW theorem to an arbitrarily small fraction of the $n$ channels
$\CM$. Doing so has no effect on the capacity region.

\medskip

\begin{IEEEproof}(converse)
\begin{figure}
\centering
\includegraphics[width=0.40\textwidth]{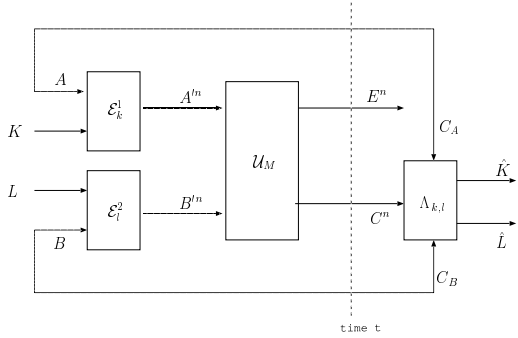}
\caption{A general protocol for multiple-access
entanglement-assisted classical communication} \label{fig:fig1}
\end{figure}
Start with some $(n,R_1,R_2,\ep)$ entanglement-assisted code (see
Fig. \ref{fig:fig1}). Assume Alice's message $k$ and Bob's message
$l$ are picked according to the uniform distributions on $[2^{n R_1
}]$ and  $[2^{n R_2 }]$, respectively. These correspond to random
variables $K$ and $L$. Alice performs the encoding operation
$\CE_k^1$ on the $A$ part of $\ket{\Gamma_1}^{AC_A}$ conditioned on
$K = k$. Bob performs the encoding operation $\CE_l^2$ on the $B$
part of $\ket{\Gamma_2}^{BC_B}$ conditioned on $L = l$. The output
of $\CE_k^1 \otimes \CE_l^2$ is  sent through the multiple access
channel ${\CM}^{\otimes n}$ just after time $t_0$ . The channel
output $C^n$ is acquired by Charlie at time $t$. Charlie performs a
POVM on the channel output and his part of the entanglement
$C_AC_B$. The measurement outcome is a random variable $W =
(\hat{K}, \hat{L})$. By the condition (\ref{macc}), \beq \label{pre}
\Pr\{ K \neq \hat{K} \, {\rm and} \, L \neq \hat{L} \} \leq
\epsilon. \eeq The protocol ends at time $t_f$. We first obtain an
upper bound on the sum rate $R_1 + R_2$.
At this time \beq \label{fano} n (R_1 + R_2) = H(KL) \leq I(KL;
\hat{K}\hat{L}) + n \eta(n, \epsilon), \eeq where the function
$\eta(n, \epsilon)$ tends to $0$ as $\epsilon$ tends to $0$ and $n$
tends to infinity. The inequality is standard in classical
information theory \cite{CT91}. It is obtained by applying  Fano's
inequality \cite{CT91} to (\ref{pre}). Denote the state of the
system at time $t$ by
$$ \omega^{KLC_A C_B C^n E^n}=(I^{KLC_A C_B} \otimes U_{\CM}^{\otimes n})
(\xi_1 \otimes \xi_2),
$$
$$
\xi_1^{A'^n  K C_A} = 2^{-n R_1} \sum_{k} \proj{k}^{K}\otimes
(\CE^1_k \otimes  I^{C_A}) (\Gamma_1^{AC_A}),
$$
$$
\xi_2^{B'^n  L C_B} = 2^{-n R_2} \sum_{l} \proj{l}^{L}\otimes
(\CE^2_l \otimes  I^{C_B}) (\Gamma_2^{BC_B}).
$$
Denote by $A^n$ the system which purifies the restriction of the
$A'^{n}$ parts of the state $\xi_1$ at time $t_0$. Then $A^n$
contains $K$ and $C_A$ as subsystems. Define $B^n$ in a similar
fashion.

The Holevo bound reads \beq I(KL; \hat{K}\hat{L}) \leq I(KL; C_A C_B
C^n)_\omega. \label{holbd} \eeq The entropic quantities below refer
to the state $\omega$ \beq
\begin{split}
& I(KL; C_A C_B C^n) \\
& = I(C^n;C_A C_B KL) - I(C_A C_B; C^n)+I(K L;C_A C_B) \\
& \leq I(C^n;C_A C_B KL) \\
& \leq  H(C^n) + H(C^n |E^n)\\
& =  H(C^n) - H(C^n |A^n B^n)\\
& =  I(C^n;A^n B^n).
\end{split}
\eeq The first inequality follows from $I(KL;C_A C_B)=0$ and $I(C_A
C_B; C^n)\geq 0$. The second inequality holds because of Lemma
\ref{lem6}. The second equality is from Lemma \ref{lem5}.

Putting everything together gives \beq R_1 + R_2 \leq \eta(n,
\epsilon) + \frac{1}{n}  I(C^n; A^n B^n) \label{ar12}. \eeq Observe
that \beq
\begin{split}
& \frac{1}{n} H(C^n) + H(C^n |E^n)\\
& \leq \frac{1}{n} \sum_i [H(C_i) + H(C_i |E_i)]\\
& \leq \max_{\rho_1, \rho_2} [H(C)_\theta + H(C|E)_\theta]\\
& = \max_{\rho_1, \rho_2} [H(C)_\theta - H(C|AB)_\theta]\\
& = \max_{\rho_1, \rho_2}  I(AB;C)_\theta.
\end{split}
\eeq The state $\theta$ is defined in (\ref{theeta}).

An upper bound on Alice's rate $R_1$ is obtained in a similar
fashion. Equations \beq
n R_1  = H(K) \leq I(K; \hat{K}) + n \eta(n, \epsilon), \eeq and
\beq I(K; \hat{K}) \leq I(K; C_A C_B C^n)_\omega
\eeq are obtained as above. With respect to $\omega$: \beq
\begin{split}
& I(K; C_A C_B C^n) \\
& = I(C_B C^n;C_A  K) - I(C_A; C_B C^n)+I(K ;C_A) \\
& \leq I(C_B C^n;C_A  K) \\
& \leq I(B^n C^n;C_A  K) \\
& \leq  H(B^n C^n) + H(B^n  C^n |E^n)\\
& = H(B^n C^n) - H(B^n  C^n |A^n)\\
& = I (A^n; B^n  C^n)\\
& = I (A^n; C^n | B^n).
\end{split}
\eeq Hence \beq R_1 \leq \eta(n, \epsilon) + \frac{1}{n} I (A^n; C^n
| B^n). \label{ar1} \eeq By the same argument \beq \label{ar2} R_2
\leq \eta(n, \epsilon) + \frac{1}{n} I (B^n; C^n | A^n). \eeq The
reason that we do not single-letterize the rates $R_1$ and $R_2$
using arguments in (\ref{ar12}) is due to the definition of systems
$A^n$ and $B^n$, which contain the classical information $K$ and $L$
as subsystems, respectively. At the same time, the channel output
$C^n$ also contains information regarding $K$ and $L$. Therefore, it
is not trivial that chain rule is applicable to systems $B^nC^n$
(likewise $A^nC^n$).

Now assume that $(R_1, R_2)$ is achievable. This means that for all
$\epsilon >0, \delta > 0$, there exists an $(n, R_1- \delta, R_2 -
\delta, \epsilon)$ code, and hence \beq
\begin{split}
R_1 &\leq \eta(n, \epsilon) + \delta + \frac{1}{n} I (A^n; C^n| B^n)\\
R_2 &\leq   \eta(n, \epsilon)  + \delta + \frac{1}{n} I (B^n; C^n | A^n) \\
R_1 + R_2 &\leq \eta(n, \epsilon) + 2 \delta + \frac{1}{n}  I(C^n;
A^n B^n).
\end{split}
\eeq It follows that $(R_1, R_2)$ is in the $\nu(n, \epsilon,
\delta)$ neighborhood of the $\frac{1}{n}{\tilde{C}}_E(\CM^{\otimes
n})$ region, with $\nu(n, \epsilon, \delta) \rightarrow 0$ as
$\epsilon \rightarrow 0, \delta \rightarrow 0, n \rightarrow
\infty$. Hence $(R_1, R_2)$ is in $C_E(\CM)$, concluding the proof
of the converse.
\end{IEEEproof}

\section{The collective phase-flip channel example}
\label{IV} Consider the case that $|A'|=|B'|=d \geq 2$. The
collective phase-flip channel \cite{Yard05}  $\CM_p:A' B' \to C$ is
defined as \beq \label{cpfc}
\CM_p(\rho)=\sum_{k=0}^{d-1}p_k(\hat{Z}(k)\otimes\hat{Z}(k))\rho
(\hat{Z}(k)\otimes\hat{Z}(k))^{\dagger} \eeq where $\hat{Z}(k)$ is
the generalized Pauli phase operator from (\ref{gpm}). We will show
that the capacity region for the multiple access phase-flip channel
$\CM_p$ assisted by entanglement is the collection of all pairs of
nonnegative rates $(R_1 ,R_2)$ which satisfy \beq \label{werwer}
\begin{split}
R_1&\leq 2 \log d \\
R_2&\leq 2 \log d \\
R_1 + R_2  &\leq  4\log d-H(p).
\end{split}
\eeq
\begin{IEEEproof} First we show that (\ref{werwer}) is
precisely the region $C_E(\CM, \pi, \pi)$, proving achievability.
The corresponding $\theta$ state is
$$
\theta^{A B C}= (I^{A B}\otimes\CM_p)(\Phi^{AA'}\otimes\Phi^{BB'})
$$
where $\ket{\Phi}$  is the maximally entangled state (\ref{maxen}).
It is easy to see that \beq
\begin{split}
H(A)&=H(B)=H(\pi)=\log d \\
H(AC)&=H(BC)=\log d +H(p) \\
H(ABC)&=H(p).
\end{split}
\eeq Hence we reach our conclusion \beq
\begin{split}
I(A;C|B)&=2 \log d \\
I(B;C|A)&=2 \log d \\
I(AB;C)&=4\log d-H(p).
\end{split}
\eeq

It remains to show that (\ref{werwer}) is an upper bound on the
capacity region. It is clear from (\ref{cr}) that  $R_1 \leq 2 H(A)$
and $R_2 \leq 2 H(B)$. Hence the first two inequalities in
(\ref{werwer}). The third makes use of the single-letter upper bound
(\ref{slub}) on $R_1 + R_2$. It suffices to show that \beq
\max_{\rho} I(A B; C)_\theta = 4\log d-H(p), \label{mahi} \eeq where
\beq \theta^{A B C}= (I^{AB} \otimes \CM) (\varphi^{AB A'B'}),
\label{theeta1} \eeq and $\varphi^{AB A'B'}$ is a purification of
$\rho^{A'B'}$. \footnote{we have already shown that this maximum is
achieved for the product state $\rho^{A'B'} = \pi^{A'} \otimes
\pi^{B'}$.} We need three ingredients. The first is that the maximum
in (\ref{mahi}) is attained for states $\rho^{A'B'}$ diagonal in the
$\{\ket{jl}\}$ basis (see Appendix \ref{gedeh} for a proof of this
fact).
Define a Stinespring dilation $U_{\CM_p}:A' B' \to C E$ of $\CM_{p}$
as \beq \label{isoext}
U_{\CM_p}=\sum_{jl}\ket{jl}^{C}\ket{\phi_{jl}}^{E}\bra{jl}^{A'B'}
\eeq where
$$ \ket{\phi_{jl}}^{E}= \sum_{k=0}^{d-1}\sqrt{p_k}\ket{k}e^{i2\pi k(j+l)/d} .$$
By the results of Appendix \ref{gedeh} \beq I(A B; C)_\theta =
2H(\{r_{jl}\})-H(\sum_{jl}r_{jl}\phi_{jl}), \eeq where $\rho =
\sum_{jl} r_{jl} \proj{jl}$.

The second ingredient is that $I(A B; C)_\theta$ is a concave
function of $\rho$ and hence has a unique local optimum.  This is
because for \emph{degradable channels} \cite{DS03} such as
$\CM_{p}$, the coherent information $I(AB\rangle C) := I(A B; C) -
H(A)$ is a concave function of input density matrix $\rho$
\cite{Yard05}. Since $H(A)$ is also concave we conclude that
$I(AB;C)$ is concave.

The third ingredient is to use the method of Lagrange multipliers to
find a local optimum for $I(A B; C)_\theta$. We need to optimize
$$
f(\{r_{jl}\}) = 2H(\{r_{jl}\})- H(\sum_{jl}r_{jl}\phi_{jl}) -
\lambda \sum_{jl} r_{jl},
$$
with Lagrange multiplier $\lambda$. Differentiating with respect to
the $r_{jl}$ gives $d^2$ simultaneous equations. By inspection,
$r_{jl} = 1/d^2$ is a solution to this system of equations. The
second ingredient ensures that this is in fact the global maximum.
Thus
\begin{align*}
\max_{\rho} I(A B; C)_\theta &= 2H(\{\frac{1}{d^2}\})-H(\frac{1}
{d^2}\sum_{m}\phi_m) \\ &= 4\log d-H(p)
\end{align*}
as claimed.
\end{IEEEproof}

\section{A hierarchy of QMAC resource inequalities}
\label{V}

In this section we phrase our result using the theory of resource
inequalities developed in \cite{DHW03}. The multiple access channel
$\CM:A' B' \rightarrow C$ assisted by some rate $E_1$ of ebits
shared between Alice and Charlie and some rate $E_2$ of ebits shared
between Bob and Charlie, was used to enable a rate $R_1$ bits of
communication between Alice and Charlie and  a rate $R_2$ bits of
communication between Bob and Charlie. This is written as
\begin{multline*}
\< \CM \> + E_1 \, [ q \, q]_{AC}  + E_2 \, [ q \, q]_{BC} \\ \geq R_1
\, [c \rightarrow c]_{AC}  + R_2 \, [c \rightarrow c]_{BC}.
\end{multline*}
Without accounting for entanglement consumption (i.e. setting $E_1 =
E_2 = \infty$) the above resource inequality holds iff $(R_1, R_2)
\in C_E(\CM)$, with $C_E(\CM)$ given by Theorem \ref{t2}. The ``if''
direction, i.e. the direct coding theorem, followed from the
``corner points'' 
\begin{multline} \< \CM \> +  H(A) \, [ q \, q]_{AC}  + H(B)
\, [ q \, q]_{BC}\\ \geq I(A;C) \, [c \rightarrow c]_{AC}  + I(B;CA)
\, [c \rightarrow c]_{BC} \label{uno} 
\end{multline} 
and 
\begin{multline} \< \CM \> + H(A)
\, [ q \, q]_{AC}  + H(B) \, [ q \, q]_{BC}\\ \geq I(A;CB) \, [c
\rightarrow c]_{AC}  + I(B;C) \, [c \rightarrow c]_{BC}. 
\end{multline} 
All the entropic quantities are defined relative to the state
$\theta^{ABC}$ defined in (\ref{theeta}).

Just as in the single user case (cf. rule O in \cite{DHW03}), the
protocol can be made coherent, replacing $[c \rightarrow c]$ by
$\frac{1}{2}([q \, q] +  [q \rightarrow q])$. Canceling terms on
both sides gives ``father'' protocols for the QMAC 
\begin{multline}
\< \CM \> +  \frac{1}{2} \, I(A; BE) \, [ q \, q]_{AC}  +  \frac{1}{2} \, I(B; E) \, [ q \, q]_{BC} \\
\geq
 \frac{1}{2} \, I(A;C) \, [q \rightarrow q]_{AC}  +  \frac{1}{2} \, I(B;CA) \, [q \rightarrow q]_{BC}
\end{multline} 
and 
\begin{multline}
\< \CM \> +  \frac{1}{2} \, I(A; E) \, [ q \, q]_{AC}  +  \frac{1}{2} \, I(B;A E) \, [ q \, q]_{BC} \\
\geq
 \frac{1}{2} \, I(A;CB ) \, [q \rightarrow q]_{AC}  +  \frac{1}{2} \, I(B;C) \, [q \rightarrow q]_{BC},
\end{multline} where the entropic quantities  are now defined with respect to
a purification $\theta^{ABCE}$ of $\theta^{ABC}$.

Applying $[q \to q] \geq [qq]$ to the above equations gives \beq \<
\CM \> \geq
 I(A \> C) \, [q \rightarrow q]_{AC}  +  \frac{1}{2} \, I(B \> CA) \, [q \rightarrow q]_{BC}
\label{qqmac1} \eeq and \beq \< \CM \> \geq
 I(A \> BC) \, [q \rightarrow q]_{AC}  +  \frac{1}{2} \, I(B \> C) \, [q \rightarrow q]_{BC}.
\label{qqmac2} \eeq These equations are of the form \beq \< \CM \>
\geq Q_1 \, [q \rightarrow q]_{AC}  + Q_2 \, [q \rightarrow q]_{BC}.
\label{qqmac} \eeq The optimal set of pairs $(Q_1, Q_2)$ satisfying
(\ref{qqmac}) was found in  \cite{Yard05}, \cite{HOW05}. Equations
(\ref{qqmac1}) and (\ref{qqmac2}) recover the ``corner points'' of
the corresponding capacity region.

Coherifying only Bob's resources in equation (\ref{uno}) gives
\begin{multline*}
\< \CM \> +  H(A) \, [ q \, q]_{AC}\\  \geq I(A;C) \, [c \rightarrow
c]_{AC} +  I(B \> CA) \, [q \rightarrow q]_{BC}.
\end{multline*}
Consider $\CM$ of a special $\{c q \rightarrow q \}$ form in which
Alice's input is dephased before being sent though the channel. The
arguments from Section \ref{ska}  apply here to show that the
Alice-Charlie entanglement is not needed. Thus we recover another
coding theorem proven in  \cite{Yard05} which characterizes the
pairs $(R_1, Q_2)$ for which
$$
\< \CM \>  \geq R_1 \, [c \rightarrow c]_{AC}  + Q_2 \, [q
\rightarrow q]_{BC}.
$$

We can also recover the result of Winter \cite{Winter01CQ} which
solves
$$
\< \CM \>  \geq R_1 \, [c \rightarrow c]_{AC}  + R_2 \, [c
\rightarrow c]_{BC}.
$$
for  $\{c c \rightarrow q \}$ channels $\CM$. We just apply the
argument from Section \ref{ska} to remove the need for any
entanglement assistance.

Ultimately we would like to solve
\begin{multline*}
\< \CM \> \geq Q_1 \, [q \rightarrow q]_{AC} +E_1 \, [ q \, q]_{AC}
+ R_1 \, [c \rightarrow c]_{AC}\\ +Q_2 \, [q \rightarrow q]_{BC} +E_2
\, [ q \, q]_{BC} + R_2 \, [c \rightarrow c]_{BC},
\end{multline*}
where the 6 rates may be positive or negative. The single user case
$Q_2 = E_2 = R_2 = 0$ was solved in \cite{DHLS05}.

\section{conclusion}
\label{VI} We derived a regularized formula for the
entanglement-assisted capacity region for quantum multiple access
channels. This expression parallels the capacity region for
classical multiple access channels. We leave it as an open problem
to single-letterize the above capacity region. We do not know if the
regularization in our main theorem is actually necessary.
Indications that it might not be are the successful
single-letterization of the two-user entanglement-assisted capacity
in \cite{BSST01} which we have used to obtain the single-letter
bound on the rate-sum above, and the fact that the regularization is
not necessary in the classical case.

Though the issue with more than 2 inputs was not addressed, we
expect it to be an easy extension. Suppose we have a QMAC $\CM$ with
$s$ senders and 1 receiver such that $\CM:A_1A_2\cdots
A_s\rightarrow B$. We conjecture the following
statement to be true \cite{Winter01CQ}:\\
{\it The entanglement-assisted capacity region of the quantum
multiple access channel $\CM$ is the regularized version of the
convex closure of all nonnegative $\{R_1,\cdots,R_s\}$ satisfying
$$\sum_{i\in J}R_i \leq I(A[J];B|A[J^c]) \ \ \forall J\subset [s],$$
where $A[J]=\{A_i|i\in [J]\}$ and
$[J^c]=[s]\backslash J$.} \\

The difficult problem would be to consider the quantum multiway
channel which has $s$ senders and $r$ receivers. We believe a
different approach might be needed.

\appendices
\section{Proof of Packing Lemma}
We need the following lemma from \cite{HN03}.
\label{Appendix:packing}

\begin{lemma}[Hayashi, Nagaoka]
\label{haynag} For any operators $0 \leq S \leq \1$ and $T \geq 0$,
we have
$$
\1 - \sqrt{S + T}^{-1} S \sqrt{S + T}^{-1} \leq 2 ( \1 - S) + 4 T.
$$
\end{lemma}
We are now ready to prove the packing lemma, along lines suggested
by the work \cite{HN03}.
\medskip

\begin{IEEEproof}
Let $X^N$ denote a sequence of random variables $X_1,X_2,\dots,X_N$,
where each random variable $X_k$ takes values from $\CS$ and is
distributed according to $\la$. Set $f(k) = X_k$. Each random code
$C=\{\sigma_{x_k}\}_{k\in[N]}$ is generated according to $X_k=x_k$.
Define $p_e(k)$ to be the probability of error for a single codeword
$\sigma_{x_k}$: $${p}_e(k)=\tr \sigma_{x_k}(I-\La_k),$$ where the
POVM elements $\{\La_k\}$ are constructed by the so-called
\emph{square root measurement}\cite{Hol98, SW97}
$$\La_k = \BL\sum_{l=1}^N \Upsilon_{x_l}\BR^{-\frac{1}{2}} \Upsilon_{x_k}
\BL\sum_{l=1}^N \Upsilon_{x_l}\BR^{-\frac{1}{2}}$$ with
$$\Upsilon_{m} = \Pi \Pi_m \Pi.$$
Define $p_e(C)$ to be the average probability of error, averaged
over all codewords in $C$:
$$p_e(C)=\frac{1}{N}\sum_{k=1}^N p_e(k).$$
Define $\bar{p}_e$ to be the average probability of error, averaged
over all possible random codes $C$ to be:
$$\bar{p}_e=\E_{X^N} \left[ p_e(C)\right].$$
The idea here is that if the average probability of error
$\bar{p}_e$ is small enough, we can then show the existence of at
least one good code. In what follows, we will first show that
$\bar{p}_e\leq \epsilon'$ for some $\epsilon' \to 0$ when $n\to
\infty.$ \\
 
Invoking Lemma \ref{haynag}, we can now place an upper bound on
$p_e(C)$:
\begin{equation} p_e(C)
\leq \frac{1}{N} \sum_{k = 1}^N \, \left[2(1-\tr\sigma_{x_k}
\Upsilon_{x_k})+4\sum_{l\neq k}\tr\sigma_{x_k}
\Upsilon_{x_l}\right]. \label{glaz}
\end{equation}
The gentle operator lemma in \cite{Winter99} and property (\ref{c1})
give
\begin{equation} \label{trdi} \| \Pi \sigma_m \Pi  - \sigma_m
\| \leq \sqrt{ 8 \epsilon}.
\end{equation}
By property (\ref{c2}) and (\ref{trdi})
\begin{eqnarray}
\tr \sigma_{m}  \Upsilon_{m} & \geq &  \tr \sigma_{m} \Pi_m -
\|  \Pi \sigma_m \Pi -  \sigma_{m} \| \nonumber  \\
& \geq &  1 - \epsilon - \sqrt{8 \epsilon} \label{trsl}.
\end{eqnarray}
For $k \neq l$, the random variables $X_k$ and $X_l$ are
independent. Thus
\begin{eqnarray}
\E_{X^N} \left[\tr \sigma_{X_k}  \Upsilon_{X_l}\right]
 & = &  \tr \left( \Pi \E \sigma_{X_k} \Pi \,\, \E \Pi_{X_l} \right) \nonumber  \\
 & \leq &  D^{-1} \E \tr \Pi \Pi_{X_l}\nonumber  \\
  & \leq &  d/D \label{dido} .
\end{eqnarray}
The first inequality follows from $\E \, \sigma_{X_k} = \sigma$ and
property (\ref{c3}). The second follows from $\Pi \leq \1$ and
property (\ref{c4}). Taking the expectation of (\ref{glaz}), and
incorporating (\ref{trsl}) and (\ref{dido}) gives
\begin{equation}
\begin{split}
\bar{p}_e & \leq 2 (\ep+  \sqrt{8 \ep}) +4 (N - 1)  d/D,  \\
&\leq 2 (\ep+  \sqrt{8 \ep}) + 4 N d/D  \\
&=2 (\ep+ \sqrt{8 \ep}) + 4 \gamma =: \epsilon'.
\end{split}
\end{equation}
Two more standard steps are needed.
\begin{enumerate}[i)]
\item Derandomization. \label{derandom}
There exists at least one particular value $x^N$ of the string $X^N$
such that this code $C^*=\{\sigma_{x^N}\}$ for which $p_e(C^*)$ is
at least as small as the expectation value. Thus
\begin{equation}
\label{fixerror} p_e(C^*)\leq \ep'.
\end{equation}
\item Average to maximal error probability. \label{maxerror}
Since $$p_e(C^*) =\frac{1}{N}\sum_{k\in N}p_e(k)\leq \ep',$$ then
$p_e(k)\leq 2 \ep'$ for at least half the indices $k$. Throw the
others away and redefine $f$, $N$ and $\gamma$ accordingly. This
further changes the error estimate to $4(\ep+\sqrt{8\ep})+8\gamma.$
\end{enumerate}
\end{IEEEproof}

\begin{remark}
The major difference between the proof of packing lemma and the
proof of HSW theorem is that the ensemble in HSW theorem is assumed
to be of the tensor power of $n$ copies of $\{\lambda_j,\rho_j\}$.
This is where the conditional typicality comes into play in order to
bound the probability of correctly identifying the classical
message. However, in packing lemma, the ensemble is assumed to be
some general states in $\CH^{\otimes n}$. Even thought the
projectors $\Pi_m$ indeed conditioned on $m$, but they are not
necessary projectors onto conditionally typical subspace, Therefore,
as we have claimed before, the proof of packing lemma only requires
typicality.
\end{remark}

\section{Proofs of properties  (\ref{pl})-(\ref{p4})}
\label{appendix1}

\begin{enumerate}[I.]
\item Proof of property (\ref{pl}). \\
Define $\check{P}$ to be the complement of the projector $P$. That
is $\check{P}= I-P$. \beq
\begin{split}
\label{C11}
\Pi&=\FTN\otimes\FT \\
&=(I-\CFTN)\otimes (I-\CFT)\\
&=I\otimes I -I \otimes\CFT-\CFTN\otimes I+\CFTN\otimes\CFT \\
&\geq I\otimes I -I \otimes\CFT -\CFTN\otimes I.
\end{split}
\eeq Therefore \beq
\begin{split}
&\tr{\S^{BB'}\Pi} \\ & \geq \tr{\S}-\tr{\S ( I \otimes\CFT )}-\tr{\S ( \CFTN\otimes I)}  \\
& = 1 - \tr[\S^{B'}\CFT ] - \tr [\S^{B} \CFTN] \\
&\geq 1-2\epsilon,
\end{split}
\eeq the last line by a double application of (\ref{q1}).
\item Proof of property (\ref{p2}).  \\
By (\ref{sigthe}) and (\ref{proje}), \beq
\begin{split}
\tr{\S\P}&= \tr \theta^\on\Pi_{\theta,\delta}^{n}\\
&\geq 1-\epsilon.
\end{split}
\eeq The last line follows from (\ref{q1}).

\item Proof of property (\ref{p3}). \\
\beq \tr{\P}=\tr{\Pi_{\theta,\delta}^n} \leq
2^{n[H({AB})_\theta+c\delta]}. \eeq The inequality follows from
(\ref{q3}).
\item  Proof of property  (\ref{p4}). \\
Because of (\ref{cc7}), we can bound the density operator $\pit$  by
\beq \label{C41}
\pit=\frac{\Pi^n_{t(\alpha)}}{\tr\Pi^n_{t(\alpha)}}\leq 2^{-n\lb
H(\rho)-\eta(\delta) \rb}\FT. \eeq Then 
\end{enumerate}
\beq
\begin{split}
&\Pi\sigma\Pi \\
&=(\FTN\otimes\FT)\lb\sum_{\alpha}\pt(\CNn(\pit)\otimes\pit)\rb(\FTN\otimes\FT)\\
&=\sum_{\alpha}\pt\lb(\FTN\CNn(\pit)\FTN) \otimes (\FT\pit\FT) \rb\\
&\leq \BL\FTN \CNn(\sum_{\alpha}\pt\pit)\FTN\BR
\otimes (2^{-n\lb H(\rho)-\eta(\delta) \rb}\FT) \\
&\leq \BL2^{-n \lb H(\CN(\rho))-c\delta\rb}\FTN\BR \otimes \BL2^{-n\lb H(\rho)-\eta(\delta)\rb}\FT\BR \\
&=2^{-n\lb H(\rho)+H(\CN(\rho))-c\delta-\eta(\delta)\rb } \, \Pi \\
&= 2^{-n\lb H(A)_\theta +H(B)_\theta -c\delta-\eta(\delta)\rb} \,
\Pi,
\end{split}
\eeq 
where the first inequality follows from (\ref{C41}) and the
second from (\ref{q2}).

\section{Generalized dephasing channels}
\label{gedeh} We follow the techniques of \cite{Yard05, Yard05a,
DS03}. Let $A'$ and $B$ be quantum systems of dimension $d$ with
respective bases  $\{\ket{i}^{A'}\}$ and $\{\ket{i}^{B}\}$.

A channel $\CN:A'\to B$ is called a generalized dephasing channel if
$$
\CN(\proj{i}^{A'})= \proj{i}^{B}.
$$
We can write down a Stinespring dilation $U_\CN:A'\to BE$ for $\CN$:
$$
U_\CN=\sum_{i}\ket{i}^{B}\ket{\phi_i}^{E}\bra{i}^{A'},
$$
where the $\{\ket{\phi_i}^{E}\}$ are not necessarily orthogonal.
Given $U_\CN$, the complementary channel $\CN^c: A' \rightarrow E =
\tr_{B} \circ U_\CN$ acts on some input state $\rho^{A'}$ as \beq
\begin{split}
\CN^{c}(\rho) & = \tr_{B}{U_\CN(\rho)}  \\
&=
\sum_{i}\bra{i}^{B}\BL\sum_{i''i'}\ket{i''}^{B}\ket{\phi_{i''}}^{E}\bra{i''}^{A'}\rho\ket{i'}^{A'}\bra{i'}^{B}
\bra{\phi_{i'}}^{E}\BR\ket{i}^{B} \\
&=\sum_{i}\bra{i}\rho\ket{i}\phi_i^E  \\
&=: \sum_{i} r_{i} \phi_i^E.
\end{split}
\eeq It depends only on the diagonal elements $\{r_i\}$ of $\rho$
expressed in the dephasing basis. When the $\{\ket{\phi_i}^{E}\}$
are also orthogonal, the channel $\CN$ is called {\em completely
dephasing} and is denoted by $\triangle$. It corresponds to
performing a projective measurement in the dephasing basis and
ignoring the result. The following properties hold \cite{Yard05a}:
\beq
\begin{split}
\CN^{c}&=\CN^{c}\circ\triangle \\
\CN\circ \triangle &= \triangle\circ \CN  \\
H(\triangle(\rho))&\geq H(\rho).
\end{split}
\eeq

Define $\theta^{AB} = (I^A \otimes \CN) \phi^{AA'}$, where
$\phi^{AA'}$ is a purification of the input state $\rho^{A'}$.

\begin{lemma}
\label{dephasing} Given a dephasing channel $\CN:A'\to B$, the
mutual information $I(A;B)_\theta$ is maximal when the input state
$\rho^{A'}$  is  diagonal in the dephasing basis.
\end{lemma}
\begin{IEEEproof} Since \beq
\begin{split}
I(A;B)&=H(A)+H(B)-H(BA) \\
&=H(A)+H(B)-H(E) \\
&=H(\rho)+H(\CN(\rho))-H(\CN^{c}(\rho))  \\
&\leq H(\triangle(\rho))+H((\triangle\circ\CN(\rho))-H(\CN^{c}\circ\triangle(\rho))  \\
&=
H(\triangle(\rho))+H(\CN\circ\triangle(\rho))-H(\CN^{c}\circ\triangle(\rho))
\end{split}
\eeq The inequality is saturated when $\rho = \triangle (\rho) =
\sum r_i \proj{i}$, in which case
$$
I(A;B)=2H(\{r_i\})-H(\sum_{i}r_i\phi_i).
$$
\end{IEEEproof}

\section*{Acknowledgment}
The authors are grateful to Jon Yard for valuable discussions on generalized
dephasing channels.

\ifCLASSOPTIONcaptionsoff
  \newpage
\fi



%
\bibliographystyle{IEEEtran}


%








\end{document}